\documentstyle[prl,aps,multicol,epsf]{revtex}
\begin{document}

\draft
\title{Spin-dependent transport in a quasi-ballistic quantum wire}

\author{C.-T.~Liang,$^{\dag}$ M.~Pepper, M.~Y.~Simmons, C.~G.~Smith and D.~A. Ritchie}

\address{Cavendish Laboratory, Madingley Road, Cambridge CB3 0HE, United Kingdom
}

\date{\today}

\maketitle

\widetext
\begin{abstract}
\leftskip 54.8pt
\rightskip 54.8pt

We describe the transport properties of a 5~$\mu$m long one-dimensional (1D) quantum wire. Reduction of
conductance plateaux due to the introduction of weakly disorder scattering are observed. In an in-plane magnetic
field, we observe spin-splitting of the reduced conductance steps. Our experimental results provide evidence that
deviation from conductance quantisation is very small for electrons with spin parallel and is about 1/3 for electrons 
with spin anti-parallel. Moreover, in a high in-plane magnetic field, a spin-polarised 1D channel shows a
plateau-like structure close to $0.3 \times e^2/h$ which strengthens with {\em increasing\/} temperatures. 
It is suggested that these results arise from the combination of disorder and the electron-electron interactions
in the 1D electron gas.

\end{abstract}

\begin{multicols}{2}
\narrowtext 

Using electron beam lithography, one is able to pattern the surface of a GaAs/AlGaAs 
heterostructure with sub-micron Schottky gates. By negatively biasing the surface 
gates, one can electrostatically squeeze \cite{Trevor} the underlying two-dimensional electron gas
(2DEG) into various shapes. The most noteworthy success of this technique is the experimental 
realisation of a one-dimensional (1D) channel -- by using a pair of \lq\lq split-gates"\cite{Trevor}, 
it is possible to define a 1D channel within a 2DEG\cite{KFB}. 
If the elastic scattering length is longer than the 1D channel length, transport through the channel
is ballistic and one 
observes conductance plateaux quantised in units of $2e^2/h$\cite{Wharam,vanWees1}.
Although 1D electron transport has been studied for more than a decade,
most experimental results can be explained within a single particle
picture without considering electron-electron interactions and spin effects 
in a 1D system. It is only more recently that a \lq\lq 0.7 structure'', 
evidence for possible spin polarisation caused by electron-electron  
interactions \cite{KJT}, has been extensively studied.
Part of the reason may be that in a clean
system the conductance is independent of the electron-electron
interactions and it is determined by entrance and exit reservoirs
\cite{Masstone}. A non-interacting 1D system should show either ballistic quantization or has a conductance
decreasing to zero with increasing length and decreasing temperature due to localization. On the other hand it has been 
suggested that the interaction is observable in the presence of backscattering \cite{Mas}, so called a dirty Luttinger
liquid, but in excess disorder may remove all semblance of 1D transport.

It is well known that as the length of a quantum wire is increased, the effect of disorder within the wire is enhanced
\cite{Nixon} and back-scattering in the channel is increased, resulting in a crossover from ballistic towards diffusive 
transport. In this paper, we present experimental results on the transport properties of a 5~$\mu$m quantum wire. In such
a regime the weak disorder within the system reduces conductance steps below quantised units in $2e^2/h$.
In an in-plane magnetic field, we observed spin-splitting of the conductance plateaux, as expected. 
Recently Kimura, Kuroki and Aoki\cite{Kimura} have proposed that in a dirty Luttinger liquid, a reduction of spin 
anti-parallel conductance occurs. That is, due to the electron-electron interactions, the conductance for
spin antiparallel electrons is smaller than that for spin parallel
electrons. We shall show that our results are consistent with their model.
When there are more than one 1D subbands occupied in the channel, our data further suggests 
that the electron transmission probability through a long quasi-ballistic channel shows oscillating 
behaviour with electron spin species.
Moreover, below the first spin-polarised conductance step, 
a plateau-like structure close to $0.3 \times e^2/h$ strengthens with {\em increasing\/} temperatures. 

The split-gate (SG) device (5~$\mu$m long and 0.8~$\mu$m wide) was lithographically defined, 300~nm above the 2DEG. 
The 2DEG has a carrier density of $3\times 10^{11}$~cm$^{-2}$ with a mobility 
of $7.5\times 10^{6}$~cm$^{2}$/Vs after brief illumination with a red light emitting diode.
Experiments were performed in a pumped $^{3}$He 
cryostat and the two-terminal conductance $G=dI/dV$ was measured using an ac excitation voltage of 
10~$\mu$V at a frequency of 77~Hz with standard phase-sensitive techniques. The in-plane magnetic field 
$B_{\parallel}$ is applied parallel to the source-drain current. To check for an out-of-plane magnetic field component, we measure
the Hall voltage. From this we know that the sample was aligned better than 0.1$^\circ$ using an in-situ rotating insert.
In all cases, a zero-split-gate-voltage series resistance due to the bulk 2DEG is subtracted from the raw data.
In the literature, it has been shown that there is an additional series resistance between a 1D channel and bulk 2DEG \cite{Timp}.
Three different samples at four cool-downs show similar behaviour and measurements taken from one of these samples are
presented in this paper.

Impurities in the spacer layer can give rise to potential fluctuations\cite{Davies} near a 1D wire.
This effect could cause nonuniformity of the quantum wire confinement potential, 
leading to a micro-constriction, a narrowest region in the channel.
To check whether conduction through a 1D channel is dominated by a micro-constriction, one can laterally
shift the conduction channel\cite{Rudi}. When the channel is moved away from a micro-constriction, the 
conductance -- split-gate voltage pinch-off characteristics would show large variations due to the sudden disappearance
of the extra confining potential due to impurities which causes the micro-constriction in the 
channel. We now show that it is {\em not} the case in our long quantum wire.
Figure 1 shows $G(V_{SG})$ when we differentially bias the two halves of the split-gates. In this case, the 1D channel
is laterally shifted by $\pm$ 57 nm \cite{Bill}. It is evident that the pinch-off voltages show a linear dependence
of the voltage difference between the two halves of the split-gate. Also resonant features and conductance plateaux 
show gradual evolution as the channel is moved laterally. This demonstrates that transport through the channel is
{\em not\/} dominated by a narrow region in the channel. As shown in Fig.~1,
resonant features superimposed upon conductance steps are clearly
observed in all traces. These resonances are believed due to potential fluctuations near the 1D channel \cite{Nixon}. 
The scattering potential within the channel does indeed vary when the 1D channel is moved laterally
as the strength of resonant features changes, as illustrated in Fig.~1. Nevertheless,                 
note that conductance plateaux deviations from their quantised values are always observed in all 11 traces.
However it is noticeable that the plateaux are reasonably intact showing that the scattering is weak and
does not vary significantly with the Fermi energy so producing a semblance of plateaux.
In this paper, we concentrate on the case where there is no potential difference between the two halves of the
SG.

Figure~2 shows conductance-split gate voltage characteristics $G(V_{SG})$ at various temperatures
$T$. With increasing $T$, the feature close to $0.8\times 2e^2/h$ and resonant features
gradually disappear. The first three conductance plateaux values increase and approach multiples of
$2e^2/h$ at the highest temperatures. In a shorter wire (3~$\mu$m), clean conductance plateaux close at multiples 
of $2e^2/h$ are observed. With increasing temperatures, the conductance plateaux become less well-defined
due to thermal smearing. Nevertheless the mid-points of the conductance 
steps remain close to multiples of $2e^2/h$ at high temperatures. This effect is not related to
the reports of decreased plateaus values. For example, recently the role of electron injection 
into V-groove quantum wires have been
studied \cite{Kaufman}. It has been shown that the observed reduction of ballistic conductance steps
is due to poor coupling between the 1D states of the wire and the 2D states of the reservoirs.
This mechanism may account for the reduced conductance plateaux observed in cleaved
edge overgrown quantum wires studied by Yacoby and co-workers 
\cite{Yacoby}, which is entirely different to the spin-dependent effects previously reported \cite{KJT}.

We now turn our attention to the reduced conductance plateaux as a function of magnetic field applied parallel to
the 2DEG $B_{\parallel}$. It is well established that a large $B_{\parallel}$ lifts the electron spin
degeneracy as first demonstrated by Wharam {\em et al.} \cite{Wharam}, causing consecutive spin-parallel 
(parallel to $B_{\parallel}$) and spin-antiparallel (anti-parallel to
$B_{\parallel}$) conductance plateaux in multiples of $e^2/h$ \cite{JTN}. Figure 3 shows
$G(V_{SG})$ at various $B_{\parallel}$. With increasing $B_{\parallel}$, the splitting of the conductance 
steps can be seen and the spin-split conductance steps values are somewhat lower than multiples of 
$e^2/h$. It is worth mentioning that the feature close to $0.8\times2e^2/h$, believed to be due to
resonant transmission through an impurity potential \cite{JTN1,CTL}, gradually disappears
with increasing $B_{\parallel}$. This result, together with the data shown in figure 1 when the
feature close to $0.8\times2e^2/h$ gradually turns into a resonant peak as the channel is laterally
shifted, show that at zero magnetic field one needs to be careful in ascribing {\em any\/} feature close 
to $0.7\times2e^2/h$ observed in a 1D channel to the \lq\lq 0.7 plateau" extensively studied by Thomas and co-workers \cite{KJT}. 
The zero split-gate voltage conductance shows a monotonic decreases with increasing $B_{\parallel}$, as illustrated in
the inset to Fig.~3. This effect is due to the diamagnetic shift of the 2DEG \cite{Weis}.

As clearly shown in Fig.~3, at $B_{\parallel} = 11$~T
the conductance does not show steps in multiples of $e^2/h$. We now use a different
view which reveals a striking behaviour.
Figure 4 now shows the difference in conductance between the mid-points of consecutive steps value $\Delta G(n) = G(n) - G(n-1)$ where $n$ is the
number of spin-split 1D subbands occupied. For $n =1$, $\Delta G(1)$ is simply the conductance step value. 
For $n \leq 6$, an oscillating behaviour is evident --  $\Delta G(n)$ approaches a
quantised value of $e^2/h$ when $n$ is an odd integer, and shows substantial deviations 
(up to 1/3) from a quantised value of $e^2/h$ when $n$ is an even integer.
For $n \geq 6$, the conductance steps are less pronounced and the striking oscillating behaviour gradually disappears.
Assuming that $\Delta G(n)$ reflects the transmission probability for the $n^{th}$ spin-split 1D subband, then
these experimental results suggest that the spin parallel electrons have almost a full transmission probability (100\%) through
the 1D channel whereas the spin anti-parallel electrons have a much lower transmission probability ($\approx$~65\%).
The semblance of the observed 1D conductance steps, together with the
observed weak resonant features in our weakly disordered 1D wire suggest that
our device is in the dirty Luttinger liquid regime. If this is
the case, then our experimental results are consistent with the model
proposed by Kimura and co-workers. Note that in their model they only consider
a two-band (spin parallel and spin antiparallel electrons) Tomonaga-Luttinger liquid.
The fact that the reduction of spin antiparallel conductance persists up to n=6
suggests that our results can be extended to a \lq\lq six-band'' (three pairs of 
spin parallel and spin antiparallel electrons) limit.

Finally we present the temperature dependence measurements which reveal even more striking behaviour.
Figure 5 shows $G(V_{SG})$ for $B_{\parallel} = 11$~T at different temperatures.
As expected, the spin-split conductance steps become less pronounced at higher temperatures due to thermal broadening.
However a plateau-like structure close to $0.3 \times e^2/h$ becomes more pronounced with {\em increasing\/}
temperatures. Also the structure approaches $0.4 \times e^2/h$ at the highest temperature.
The reason for this unexpected behaviour is not fully understood at present
but we speculate that 
the strong electron-electron interactions might play a role in this.

The oscillating, spin dependent transmission probability which we observed is in striking contrast to the behaviour
of a short, \lq\lq point contact'' ballistic channel where the quantization is always in units of $e^2/h$ regardless
of spin orientation. In the latter case, theory shows that due to mixing of the reservoir and channel states there can
be no electron-electron interaction enhanced deviation from the quantized values. However Kimura and co-workers
\cite{Kimura} have consider the Tomonaga-Luttinger liquid when backscattering occurs, they find that this mixes with the 
interaction to produce a conductance dependent on the Fermi energy and hence spin state although it is still
surprising the spin parallel electrons display quantized value. Recently it has been suggested that
in the Tomonaga-Luttinger regime charge density wave formation can give rise to a fractional charge behaviour\cite{Pono}.
The results here are consistent with the spin-antiparallel electrons showing a quantized conductance but with
the value of the fundamental charge reduced. We note that we cannot attribute the spin-dependent behaviour
to any spin-dependent scattering at the entrance and the exit to the channel as it is absent on short devices on the same
heterostructure material and the effect reproducible despite a change in sample and channel location.

In summary, we have performed low-temperature measurements on a quasi-ballistic quantum wire.
Our results suggest that the electron transmission probability 
through a long quasi-ballistic channel shows oscillating behaviour with spin species.
Moreover, a spin-polarised 1D channel shows a pronounced plateau-like structure close to $0.3 \times e^2/h$ 
with {\em increasing strength\/} at higher temperatures. 
Such striking behaviour is only observed in long quantum wires ($\approx$ 5~$\mu$m) but not in
a clean 1D channel ($\leq$ 3~$\mu$m), suggesting that the back-scattering within the quasi-ballistic 1D system 
plays an important role. 

This work was funded by the UK EPSRC, and in part, by the US Army Research Office. 
We thank C.J.B.~Ford for helpful discussions, K.J. Thomas for drawing our attention to Ref.~\cite{Kimura}, 
H.D.~Clark, J.E.F.~Frost and M.~Kataoka for advice and 
help on device fabrication at an early stage of this work, and S.~Shapira for experimental assistance.  
C.T.L. is grateful for support from Department of Physics, National Taiwan University.

\centerline{Figure Captions}

Figure 1.
$G(V_{SG})$ measurements when the channel is laterally shifted by
differentially biasing the two halves of the split-gates. From left to right: $\Delta V_{SG}$
= -0.5~V to +0.5~V in 0.1~V steps. The measurement temperature
was 0.3~K.

Figure 2.
$G(V_{SG})$ at various temperatures $T$ as illustrated in the figure.

Figure 3.
$G(V_{SG})$ at various applied magnetic field parallel to the 2DEG $B_{\parallel}$.
From left to right: $B_{\parallel}$ = 0 to 11~T in 1~T steps. Curves are successively
offset by 0.01~V for clarity. The zero split-gate voltage conductance at various $B_{\parallel}$, 
as shown in the inset to Fig.~3 has been subtracted from the raw data. The inset shows the 
zero split-gate voltage conductance as a function of $B_{\parallel}$. The measurement temperature
was 0.3~K.

Figure 4.
The difference in conductance between the $(n-1)^{th}$ and the $n^{th}$ 1D conductance steps
$\Delta G(n) = G(n) - G(n-1)$, where $n$ is the spin-split subband index.

Figure 5.
$G(V_{SG})$ for $B_{\parallel}$ at various temperatures $T$.
From left to right: $T =$ 0.3, 0.363, 0.391, 0.429, 0.483, 0.542, 0.611, 0.680, 0.752, 0.832,
0.920, 1.01, 1.07, 1.16, 1.28, 1.46 and 1.60~K. Curves are successively
offset by 0.01~V for clarity. The data was taken at the second cool-down.

$^{\dag}$Present address: Department of Physics, National Taiwan University,
Taipei 106, Taiwan

\end{multicols}

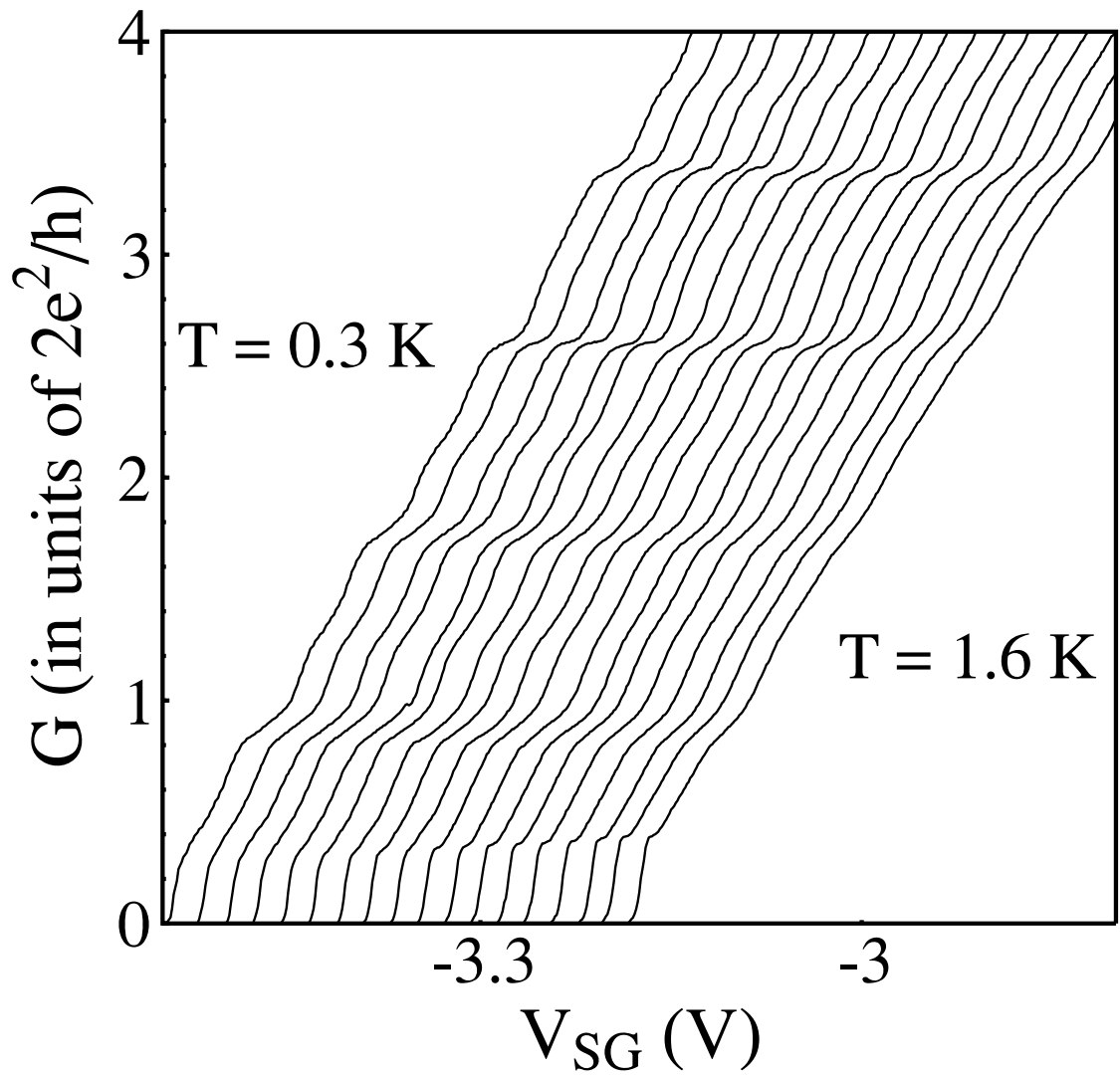
\end{document}